# Improvement of Indoor Radio Coverage at 60 GHz in NLOS Configuration

Mbissane Dieng, Marwan El Hajj, Gheorghe Zaharia, Ghais El Zein
*Univ Rennes, INSA Rennes, CNRS, IETR, UMR 6164, F 35000, Rennes, France*
Mbissane.Dieng@insa-rennes.fr

*Abstract*—In the current development of new technologies, the world of communications is experiencing significant growth thanks to the integration of wireless communications in millimeter band. In this context, the purpose of this paper is to assess the contribution that the use of a metallic reflector panel can introduce to indoor radio coverage at 60 GHz in an NLOS (corridor) configuration. The results obtained by measurement show that the use of such a panel can lead to a significant path loss reduction up to 12 dB, which improves the reception power.

## I. INTRODUCTION

The last few years have been marked by a constant evolution of wireless technologies such as WiFi (802.11ad / ay), mobile networks (from 4G to 5G) or networks of connected objects [1]. This development has allowed users to access many types of applications such as video streaming, telephony, high definition television (HDTV), fast transfer of large files, etc. These applications require very high data rates and sufficient bandwidth to meet a high demand from users faced with a limited spectral resource. In this context, the use of the millimeter wave (mmW) frequencies, which cover the 30 to 300 GHz band, ensures very high data rates and increases the capacity of networks [2].

In fact, compared to the bands below 6 GHz, used by current wireless networks, mmW bands offer advantages in terms of higher bandwidths. Moreover, corresponding shorter wavelengths allow the implementation of massive MIMO and adaptive beamforming techniques. However, mmW signals are subject to higher propagation losses and penetration attenuation through materials and and that's why they are mainly dedicated to line-of-sight (LOS) transmissions. Recent research have revealed that these signals can take advantage of these challenges, employing adaptive beamforming techniques and using relay stations or passive reflectors to bypass obstacles thus avoiding obstructions [3]-[5].

Based on measurement results, the aim of this article concerns the evaluation of power losses as a function of distance, with emphasis on the effect of introducing a reflective metal panel on the propagation of waves at 60 GHz, in the case of non-line-of-sight (NLOS).

The rest of this paper is developed as follows: Section II describes the measurement environment and the 60 GHz measurement system. Section III presents the obtained results. Finally, in Section IV, a conclusion is drawn.

## II. INDOOR PROPAGATION MEASUREMENT CAMPAIGN

This section presents the measurement campaign carried out at 60 GHz in a corridor at the IETR laboratory at INSA Rennes.

### A. Measurement Environment

The measurements were carried out in an L-shaped corridor (Fig. 1), which consists of two parts: part A has the dimensions of $3.69 \times 2 \times 3$ m$^3$ and part B is $4.7 \times 1.62 \times 3$ m$^3$. The environment is assumed to be static, without the presence of people or any other movements.

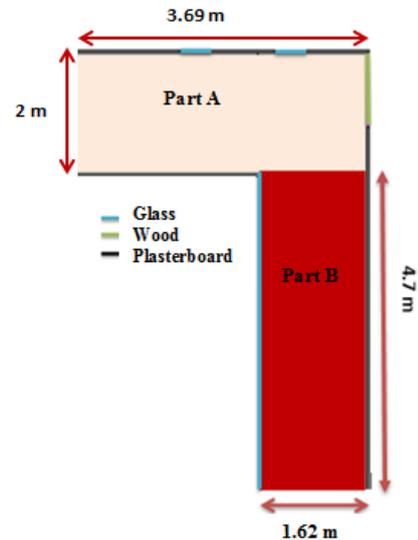

Fig. 1. Measurement environment.

The walls of the measurement environment have been made by plasterboard. In part A, we note the presence of two glass doors and a wooden door. In part B, one side of the corridor is completely glazed. Fig.2 illustrates two photos referring to each part of the hallway.

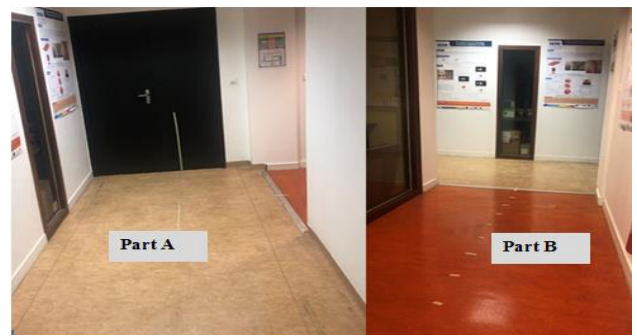

Fig. 2. View of the measurement environment.

## B. 60 GHz Measurement System

The block diagram of the complete measurement system is shown in Fig. 3.

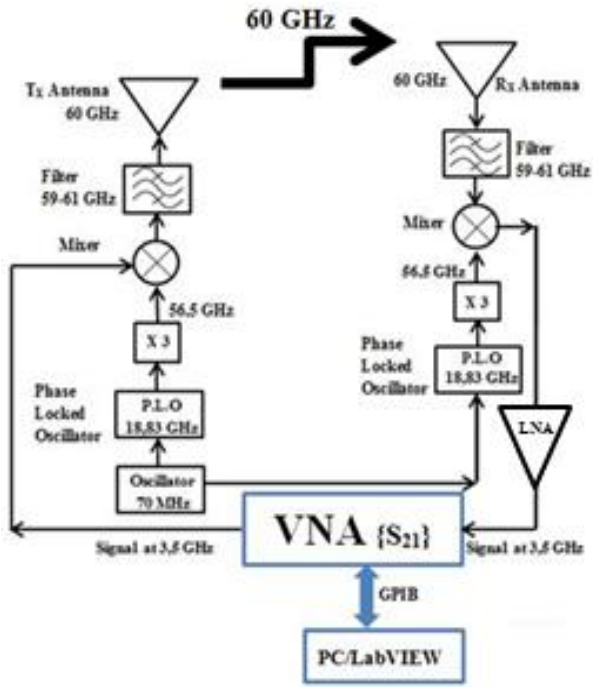

Fig. 3. 60 GHz measurement system [6]

The measurement system used is built around a vector network analyzer (VNA) which performs a frequency sweep over a 2 GHz band centered on the intermediate frequency (IF) of 3.5 GHz. The VNA, linked to the transmitter (Tx) and receiver (Rx) blocks respectively via port 1 and port 2, allows to measure the $S_{21}$ parameter which allows to calculate the frequency response of the propagation channel. The VNA scans the selected 2 GHz frequency band on 401 points with a frequency step of 5 MHz. The VNA is controlled by a computer via a GPIB interface which makes it possible to save the measurement data using a LabVIEW program developed at the IETR [6].

The Tx and Rx blocks are up and down converters allowing to transpose the frequency from 3.5 GHz to 60 GHz and inversely. The block diagram of the complete measurement system, shown in Fig. 3, includes:

- A phase locked oscillator (PLO) at 18.83 GHz on an external signal at 70 MHz
- A tripler providing the local oscillator at 56.5 GHz (exactly 56.49 GHz)
- Up converter and down-converter mixers at Tx and Rx respectively
- RF bandpass filters used to limit the signal band to 2 GHz
- A low noise amplifier (LNA), with a gain of 46 dB which is used at 3.5 GHz in reception to compensate the strong propagation losses.

A photo of the RF devices is given in Fig. 4.

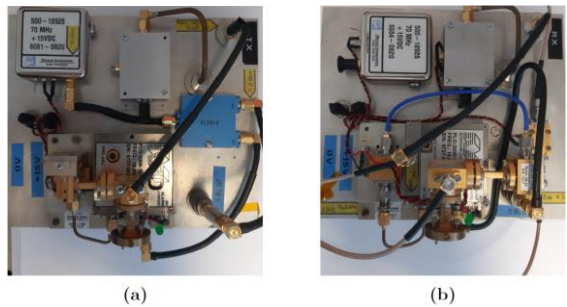

Fig. 4. The RF modules at transmission (a) and at reception (b).

Two types of antennas are used in these measurements (Fig. 5): an omnidirectional antenna is used at Tx side, with a gain of 2 dBi in azimuth and an aperture of 30° at -3 dB beam width (HPBW) in the elevation plan. At Rx, a horn antenna is used, with a 22.5 dBi gain and a -3 dB beam width of 13° in azimuth and 10° in elevation [7].

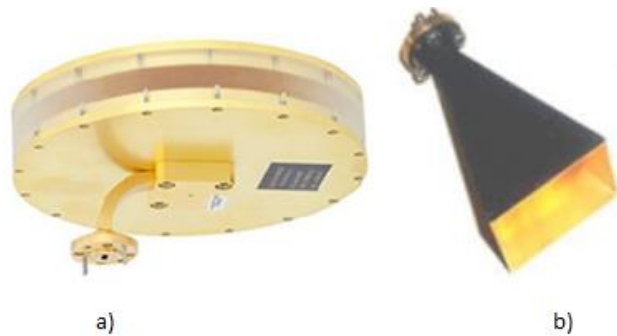

Fig. 5. Omnidirectional antenna a) and horn b) at 60 GHz.

Table I summarizes the parameters used for our measurement system.

TABLE I. MEASUREMENT PARAMETERS AT 60 GHZ

| | |
|---|---|
| Center frequency (GHz) | 60 |
| Intermediate frequency (GHz) | 3.5 |
| Bandwidth (GHz) | 2 |
| Number of points | 401 |
| Frequency step (MHz) | 5 |
| IF transmit power (dBm) | 10 |
| RF transmit power (dBm) | 0 |

## C. Channel Frequency Response at 60 GHz

At the beginning of the measurement campaign, losses were removed from the cables that connect the two RF heads to the VNA ports.

As shown in Fig. 6, $H_M$ represents the frequency response measured by the VNA, which can be expressed by:

$$H_M = H_{T_x}.H_{G_{T_x}}.H_C.H_{G_{R_x}}.H_{R_x} \quad (1)$$

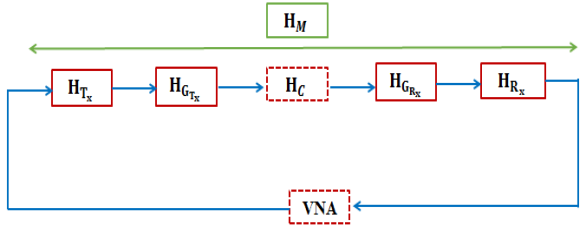

Fig. 6. The measured frequency response.

$H_{Tx}$ and $H_{Rx}$ are the frequency responses of the RF modules at Tx and Rx respectively, while $H_{G_{Tx}}$ and $H_{G_{Rx}}$ represent their antenna gains.

From (1), we can derive the expression of the channel frequency response by:

$$H_C = \frac{H_M}{H_{Tx} H_{G_{Tx}} H_{G_{Rx}} H_{Rx}} \quad (2)$$

A back-to-back test will be necessary to remove the losses introduced by the two RF modules, in order to eliminate their effects on the measured frequency response. For this, a 40 dB attenuator is introduced between the transmitter and the receiver [6] (Fig. 7).

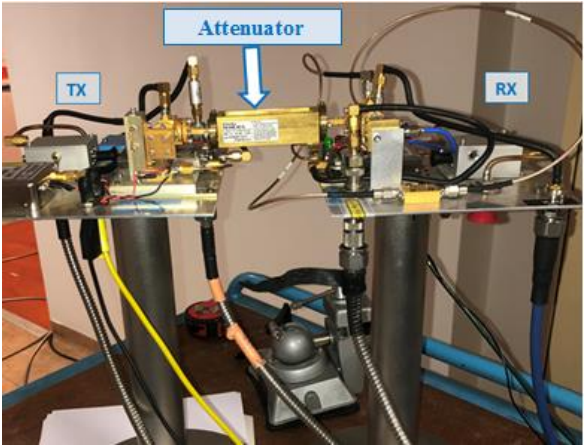

Fig. 7. Back-to-back measurement at 60 GHz.

The response measured in back-to-back is given by:

$$H_{BB} = H_{Tx} H_A H_{Rx} \quad (3)$$

Using (3), the frequency response of the channel becomes:

$$H_C = \frac{H_M H_A}{H_{BB} H_{G_{Tx}} H_{G_{Rx}}} \quad (4)$$

*D. Measurement Scenario*

During this measurement campaign, the Rx horn antenna was placed in a fixed position in the middle of the width the corridor (part A), at a distance of $L_R$ (3.69 m) from the bottom of the corridor (Fig. 8). While the omnidirectional transmitting antenna has been moved to 16 positions ranging from Tx1 to Tx16, with a step of 25 cm, along the central axis of part B of the corridor, as shown in Fig. 8.

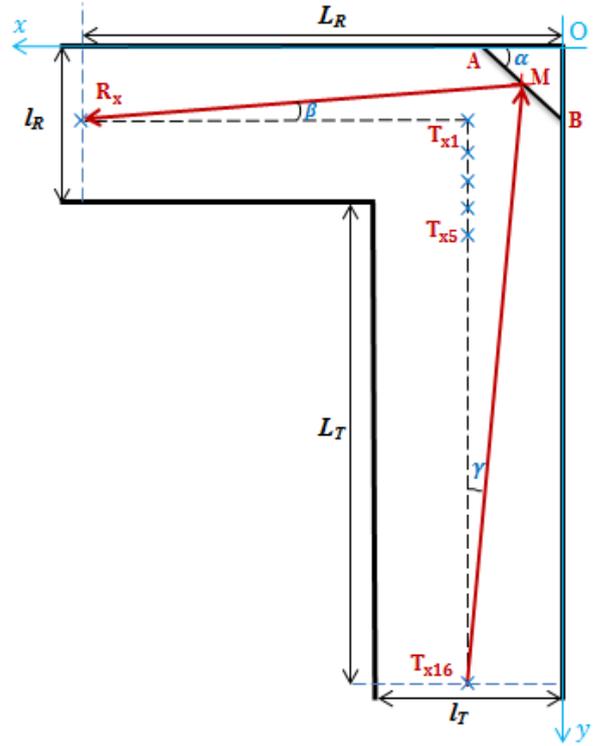

Fig. 8. Measurement scenario.

A metal plate $L_p$ = 98.2 cm length and $l_p$ = 59.5 cm width was placed in the corner of the corridor on a wooden table. The width of the panel was horizontal and its length vertical. Its center M was at a height of h = 1.37 m from the ground, the same height as for the Tx and Rx antennas.

In order to favor the farthest position Tx16, with a path directed to the center M of the panel and forming an angle $\gamma$ with the corridor axis, the metal panel is oriented at an angle $\alpha$ with respect to the corridor length, as shown in Fig. 8. The horn Rx antenna was oriented at an angle $\beta$, in order to be properly oriented towards the center M of the metal panel. The incident angle of the path between Tx and M is $\alpha+\beta$.

In order to compute the optimum value of the angle $\alpha$, we consider an orthogonal system of axes xOy. Using this system, the main points have the coordinates: $M(\frac{a}{2}\cos\alpha, \frac{a}{2}\sin\alpha)$, $T(\frac{l_T}{2}, L_T + l_R)$ and $R(L_R, \frac{l_R}{2})$. The angles between $T_{x16}M$ and Ox are $2\alpha+\beta$ and $\frac{\pi}{2} - \gamma$, so:

$$(2\alpha + \beta) + \left(\frac{\pi}{2} - \gamma\right) = \pi$$

We obtain:

$$2\alpha + \beta - \gamma = \frac{\pi}{2} \quad (5)$$

Using right triangles, we can calculate:

$$\tan\beta = \frac{\frac{l_R}{2} - \frac{a}{2}\sin\alpha}{L_R - \frac{a}{2}\cos\alpha} \quad (6)$$

and

$$\tan \gamma = \frac{\frac{l_T}{2} - \frac{a}{2}\cos\alpha}{L_T + l_R - \frac{a}{2}\sin\alpha} \quad (7)$$

Using (6) and (7), (5) becomes:

$$2\alpha + \arctan\frac{\frac{l_R}{2} - \frac{a}{2}\sin\alpha}{L_R - \frac{a}{2}\cos\alpha} - \arctan\frac{\frac{l_T}{2} - \frac{a}{2}\cos\alpha}{L_T + l_R - \frac{a}{2}\sin\alpha} = \frac{\pi}{2} \quad (8)$$

For $a = l_p = 59.5$ cm, $L_T = 2.75$ m, $L_R = 3.69$ m, $l_T = 1.62$ m and $l_R = 2$ m, we obtain by solving this non-linear equation $\alpha = 42.198°$, $\beta = 12.987°$ and $\gamma = 7.383°$, the coordinates of A and B are respectively $x_A = 44.08$ cm, $y_B = 39.97$ cm. These values where used for an optimal orientation of the vertical metal plate and the horn $R_x$ antenna.

### III. MEASUREMENT RESULTS

Based on these measurements, the results obtained show the impact of the reflective metal panel on the propagation signals at 60 GHz in the NLOS configuration.

Fig. 9, represents the evolution of the path loss in function of the different Tx positions, with and without the metal reflector.

Globally, results show that the path loss increases by moving the transmitter from position Tx1 to Tx16.

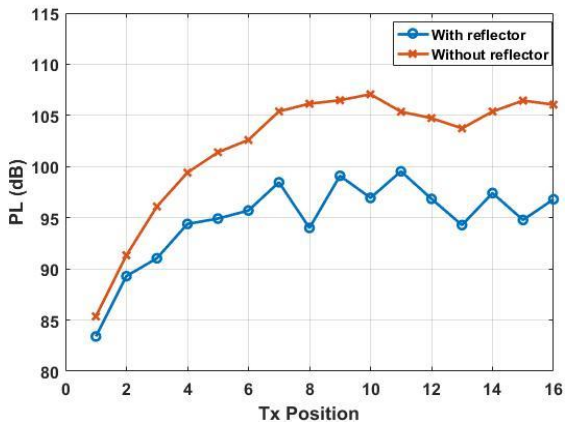

Fig. 9. Power losses for 16 Tx locations

In the presence of the metal reflector, we remark that the path loss decreases up to reach a maximum of 12 dB at position 8.

However, without the presence of the reflector, from position 8 the path loss get close to the noise floor of the VNA, which is about 108 dB, when we arrived at position 8. Hence, from the most distant position, we note a PL about 10 dB lower, which allows us to extend radio coverage at 60 GHz in NLOS zones.

### IV. CONCLUSION

This article presented the results of indoor propagation measurements carried out at 60 GHz in an L-shaped corridor. Path loss measurements have been conducted with and without reflector. The analysis of the results shows that the path loss can decrease significantly up to 12 dB which improve the radio coverage in the non-line of sight configuration.

#### ACKNOWLEDGMENT

This work was carried out as part of the ANR MESANGES project.